\documentclass[aps,prb,superscriptaddress,twocolumn,showpacs,longbibliography]{revtex4-1}
\usepackage{amsmath}
\usepackage{amsthm, amssymb} 
\usepackage{bbold}                       
\usepackage{graphicx}
\usepackage{color}
\usepackage{times, verbatim}      

\usepackage{hyperref}
\hypersetup{       colorlinks=true, }

\begin{document}

\title{Abelian and non-Abelian states in $\nu=2/3$ bilayer fractional quantum Hall systems}

\author{Michael R. Peterson}
\affiliation{Department of Physics \& Astronomy, California State University Long Beach,  Long Beach, California 90840, USA}
\author{Yang-Le Wu}
\affiliation{Joint Quantum Institute and Condensed Matter Theory
  Center, Department of Physics, University of Maryland, College Park, MD 20742}
\author{Meng Cheng}
\affiliation{Station Q, Microsoft Research, Santa Barbara, California 93106-6105, USA}
\author{Maissam Barkeshli}
\affiliation{Station Q, Microsoft Research, Santa Barbara, California 93106-6105, USA}
\author{Zhenghan Wang}
\affiliation{Station Q, Microsoft Research, Santa Barbara, California 93106-6105, USA}
\affiliation{Department of Mathematics, University of California, Santa Barbara, California 93106, USA}
\author{Sankar Das Sarma}
\affiliation{Joint Quantum Institute and Condensed Matter Theory
  Center, Department of Physics, University of Maryland, College Park, MD 20742}

\begin{abstract}
There are several possible theoretically allowed non-Abelian fractional quantum Hall (FQH) states
that could potentially be realized in one- and two- component FQH systems at total filling fraction $\nu = n+ 2/3$, for integer $n$. 
Some of these states even possess quasiparticles with non-Abelian statistics that are powerful enough for
universal topological quantum computation, and are thus of particular interest. Here, we initiate a systematic numerical
study, using both exact diagonalization and variational Monte Carlo, to investigate the phase diagram of FQH systems
at total filling fraction $\nu = n+2/3$, including in particular the possibility of the non-Abelian $Z_4$ parafermion state. 
In $\nu = 2/3$ bilayers, we determine the phase diagram as a function of interlayer tunneling and repulsion, finding only three competing
Abelian states, without the $Z_4$ state. On the other hand, in single-component systems at $\nu = 8/3$, we find that 
the $Z_4$ parafermion state has significantly higher overlap with the exact ground state than the Laughlin state, together with a 
larger gap, suggesting that the experimentally observed $\nu = 8/3$ state 
may be non-Abelian. Our results from the two complementary numerical techniques agree well with each other qualitatively.
\end{abstract}

\date{\today}


\maketitle

\section{Introduction}
Multi-component fractional quantum Hall (FQH) states appear in a wide 
variety of two-dimensional electron systems (2DES)~\cite{dasSarma1996}, such as  
multilayer or multisubband quantum wells~\cite{liu2015},  systems with small Zeeman energy where the electron spin plays an active role, and 
 systems with multiple valley degrees of freedom, such as graphene~\cite{du2009,bolotin2009,dean2011,feldman2012,feldman2013}, 
silicon~\cite{kott2014}, and AlAs~\cite{bishop2007,padmanabhan2009}. These systems  offer several tunable parameters, 
which allow  the observation of rich zero temperature phase diagrams involving  
topologically distinct FQH states even at a fixed total filling fraction, and indeed novel FQH phases of 
multicomponent systems have been  experimentally observed. Most notable perhaps 
is the observation of the so-called 331 Abelian even-denominator FQH state in half-filled 
bilayer systems~\cite{eisenstein1992,suen1992,he1993}.  
However, in many cases, little is known about the myriad possible FQH phases and phase transitions 
that can be experimentally realized in multi-component 2DES. 

Recently, motivated by the possibility of a non-Abelian state at $\nu=5/2$ in GaAs quantum wells~\cite{Moore1991,willett1987},
there have been detailed numerical studies at total filling fraction $\nu = n+1/2$ ($n$ integer) 
in two-component systems \cite{peterson2010}.
While $\nu = n + 1/2$ has been studied in great detail, the  problem at $\nu = n+2/3$ has received comparatively less attention from numerical studies~\cite{MorfPRL1998,TokePRB2005,PetersonPRB2008,WojsPRB2009,FriedmanIJMPB2010,BiddlePRB2011,DavenportPRB2012,BalramPRL2013}.
Such systems were first studied experimentally over 20 years ago, 
where a two-component to single-component phase transition was observed in  monolayer (presumably due to spin) and
bilayer systems~\cite{eisenstein1990,suen1994,manoharan1997,lay1997}. 
There are three  Abelian FQH states that can be realized at $\nu =2/3$: the $330$ state, 
\begin{equation}
\Psi_{330} = \prod_{i<j}(z_i-z_j)^3  (w_i-w_j)^3 \prod_{i,j} (z_i - w_j)^0,
\end{equation}
consisting of two decoupled $1/3$ Laughlin states in each layer where $z_i$ and $w_i$, for $i = 1,\ldots,\frac{1}{2}N$ are the complex coordinates of the electrons in the two layers, and here and hereafter we have omitted the Gaussian factor $\exp(-\sum_i |z_i|^2/4l^2)$ for all wave functions, a pseudo-spin singlet Abelian state, here called the $112$ state, 
\begin{equation}
\Psi_{\text{singlet}} = \mathcal{P}_{\text{LLL}} \prod_{i<j} |z_i - z_j|^2 |w_i - w_j|^2 \Psi_{112}^*,
\end{equation}
where $\mathcal{P}_{\text{LLL}}$ is the LLL projection operator, 
and the particle-hole conjugate of the $1/3$ Laughlin state, referred to here as the $2/3$ Laughlin state,
\begin{equation}
\Psi_{\text{P-H}} = \mathcal{P}_{\text{LLL}} \prod_{i<j}(z_i - z_j)^2 \Phi_{\nu = -2},
\end{equation}
where $\Phi_{\nu=-2}$ is the wave function for the $\nu=-2$ integer quantum 
Hall state.   The pseudo-spin singlet $112$ state can be easily understood 
within composite fermion theory as composite fermions filling the lowest spin-up and spin-down levels in a reversed 
effective magnetic field~\cite{WuPRL1993,jainCF,DavenportPRB2012}.
Early numerical work on $\nu=2/3$ bilayers
considered the overlap of model wave functions  with the exact ground state of the system
for $N = 6$ electrons on a torus~\cite{mcdonald1996}, finding these three phases in the 
two-component 2D system (for the monolayer spinful system, the $330$ state is unlikely). 

\begin{table}[t]
\caption{
\label{statesTable}
Candidate Abelian and non-Abelian FQH states at total filling fraction $\nu = 2/3$. On the sphere, these states  occur at different shifts $\mathcal{S} \equiv  \frac{3}{2} N - N_\Phi $, where $N_\Phi$ is the number of flux quanta. The Fibonacci state, as a single-component system, has a shift of 6; as a two-component system, it  has a shift of 3 per layer. 
}
\begin{ruledtabular}
\begin{tabular}{lcc}
 Possible States at $\nu = 2/3$  & Type & Shift, $\mathcal{S} $ \\
\hline
 330 State & Abelian & 3 \\
 Pseudo-Spin Singlet 112 & -- & 1 \\
 Particle-Hole Conjugate of 1/3-Laughlin & -- & 0 \\
 \hline
 $Z_4$ Parafermion & Non-Abelian & 3 \\
 Bilayer Fibonacci & -- & 3 \\
 Interlayer Pfaffian & -- & 3 \\
 Intralayer Pfaffian & -- & 3 \\
 Bonderson-Slingerland Hierarchy & -- & 4 \\
\end{tabular}
\end{ruledtabular}
\end{table}

Different theoretical studies have suggested five possible exotic non-Abelian FQH states  can occur at 
$\nu=2/3$, yet have not been numerically investigated (see Table~\ref{statesTable}). These include: the $Z_4$ parafermion
FQH state \cite{read1999,rezayi2010,barkeshli2010}, 
\begin{equation}
\Psi_{Z_4}=\mathcal{A}[\Psi_{330}],
\label{eq:z4}
\end{equation}
where $\mathcal{A}$ is an anti-symmetrization over all electron coordinates, a Fibonacci state based on $SU(3)_2$ Chern-Simons theory~\cite{wen1991prl,vaezi2014}, 
interlayer and intralayer Pfaffian states~\cite{ardonne2002,barkeshli2010multi}, 
\begin{equation}
\Psi_\text{Inter Pf}=
\text{Pf}\left(\frac{1}{z_i-z_j}\right)\,
\text{Pf}\left(\frac{1}{w_i-w_j}\right)\,\Psi_{221},
\end{equation}
and
\begin{equation}
\Psi_\text{Inter Pf}=
\text{Pf}\left(\frac{1}{x_i-x_j}\right)\,\Psi_{221},
\end{equation}
respectively, with $x_i$ running over the coordinates of the $N$ electrons in both layers with 
\begin{eqnarray}
\Psi_{221} = \prod_{i<j}(z_i-z_j)^2  (w_i-w_j)^2 \prod_{i,j} (z_i - w_j)^1,
\end{eqnarray}
and a Bonderson-Slinglerland hierarchy state~\cite{bonderson2008}. 
It should be noted that the $Z_4$ parafermion and the intralayer Pfaffian 
states are defined only when the number of electrons $N$ is divisible by $4$.  

As a result, the $N=6$ overlap study~\cite{mcdonald1996} did not actually rule 
out the possibility of having stable non-Abelian phases even in the lowest Landau level.
The $Z_4$ and Fibonacci states have  been shown theoretically to 
exhibit continuous phase transitions from the $330$ state~\cite{barkeshli2010prl,vaezi2014}, suggesting  these  states might be 
stabilized nearby more conventional ones if  appropriate microscopic parameters are found and tuned experimentally. 
The goal of our work is to investigate numerically the possible existence of exotic non-Abelian 2/3 
(or  generally, $n+2/3$) FQH states in realistic 2DES.

The Fibonacci FQH state contains the non-Abelian Fibonacci quasiparticle, whose braiding statistics is known to be powerful
enough to be utilized for universal topological quantum computation (TQC)~\cite{freedman2002}. The $Z_4$ parafermion FQH state
is based on the $SU(2)_4$ topological quantum field theory, which  has been discovered to  allow for universal TQC~\cite{cui2014,LevaillantArxiv2015}. The Bonderson-Slingerland hierarchy state at $\nu = n+2/3$, and the interlayer Pfaffian state,
can also be used for universal TQC if  realized on topologically non-trivial spaces with  topological operations 
known as Dehn twists~\cite{bravyi200universal,bondersonpc} that can be  realized in a 
physically realistic experimental setup~\cite{barkeshli2013genon,barkeshli2014qubit}. It is thus timely to revisit the 
$\nu =2/3$ bilayer phase diagram numerically and investigate the possibility of realizing these  non-Abelian states. 

In this work, we carry out a study of two-component FQH systems at total 
filling fraction $\nu=n+2/3$. We analyze the relative stability of the three 
Abelian states and the non-Abelian $Z_4$ state through exact diagonalization 
and variational Monte Carlo studies that also consider  the 
inter/intralayer Pfaffian states.
In the lowest Landau level (LLL), our results are consistent with the phase diagram proposed previously~\cite{mcdonald1996} and we importantly find that the $Z_4$ state is not competitive relative to the other Abelian states. However, in the limit of large interlayer tunneling
in the second Landau level (SLL), at $\nu = 8/3$, our  results suggest that the $Z_4$ state is preferable relative to the possible Abelian states. 
This unexpected new finding 
suggests the already experimentally observed $8/3$ FQH state may be the exotic $Z_4$ 
non-Abelian state, rather than the usual Abelian Laughlin state.  
Given the existence of the $5/2$ FQH state in the SLL,  thought to be the non-Abelian Moore-Read state, the possibility that the SLL $8/3$ FQH state might also be a (different) non-Abelian state is plausible and consistent with the fact that the experimental $8/3$ state typically is considerably weaker than the 5/2 state as manifested in the measured
energy gaps~\cite{PanPRL1999,XiaPRL2004,MillerNaturePhys2007,PanPRB2008,dean2008,choi2008,RaduScience2008,KumarPRL2010,SamkharadzePRB2011,PanPRL2012,BaerPRB2014,Kleinbaum2015}.

We consider the Hamiltonian describing two quantum Hall layers with $N$ total 
spin-polarized electrons, separated by a distance $d$, with interlayer 
electron tunneling strength $\Delta$:
\begin{equation}
\label{Hbilayer}
\!\!H\!=\!\sum_{i<j}^N
\!\Big[\!
\sum_\sigma^{\uparrow,\downarrow}\!
V_\text{intra}(|\mathbf{r}^\sigma_i-\mathbf{r}^\sigma_j|)
+V_\text{inter}(|\mathbf{r}^\uparrow_i-\mathbf{r}^\downarrow_j|)
\Big]
-\frac{e^2}{\epsilon l} \Delta S_x,
\end{equation}
where $\mathbf{r}^\sigma_i$ is the position of the $i^\text{th}$ electron in 
layer $\sigma$, and $l$ is the magnetic length.
The intralayer Coulomb interaction is given by $V_{\text{intra}}(r) = \frac{e^2}{\epsilon r}$, while the interlayer interaction is given by 
$V_\mathrm{inter}(r) = \frac{e^2}{\epsilon} \frac{1}{\sqrt{r^2+d^2}}$ ($\epsilon$ is the dielectric of the host semiconductor).  
The interlayer tunneling term is written as the total pseudo-spin $S_x$ operator, with $\Delta$ the interlayer tunneling strength in units of $\frac{e^2}{\epsilon l}$.

\begin{figure}[t]
\begin{center}
\includegraphics[width=8.5cm]{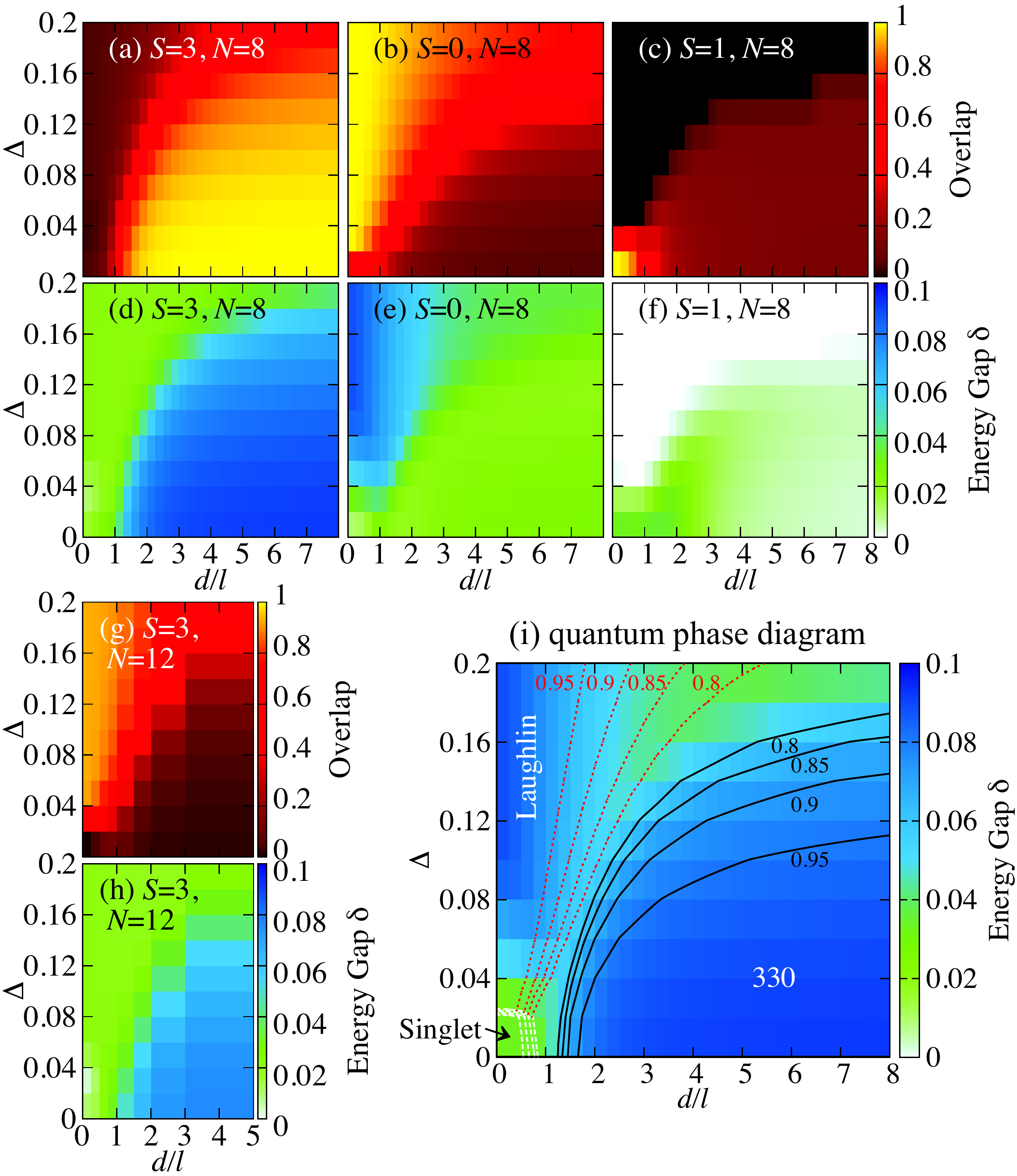}
\caption{Panels (a)-(f) correspond to shifts of $\mathcal{S}=3$, 0, and 1 from left to right for $N=8$ electrons. (a) - (c) show the overlap with $\Psi_{330}$, 
$2/3$ Laughlin, and the pseudo-spin singlet (112), respectively. (d) -(f) show the energy gap. (g) shows the overlap with $Z_4$ state for $N = 12$ at $\mathcal{S} = 3$. (h) shows the gap 
for $N = 12$ and shift $\mathcal{S} = 3$. Panel (i) displays the resulting quantum phase diagram. 
 }  
\label{fig:over-gap-N8-LLL}
\end{center}
\end{figure}

\section{Phase diagram in the lowest Landau level}
We first consider the $\nu=2/3$ bilayer quantum phase diagram  in  the LLL. The Hamiltonian 
Eq.~(\ref{Hbilayer}) has two  dimensionless parameters: $d/l$, the ratio of the inter and intralayer Coulomb interactions,
and $\Delta$,  the ratio of the interlayer tunneling to the intralayer Coulomb interaction. 
The relative stability of the three Abelian states was studied through wave function overlap calculations  for 
$N = 6$ electrons ($3$ per layer) on the torus~\cite{mcdonald1996}. We revisit this for larger systems using exact diagonalization (ED)
for up to $N = 12$ electrons in the spherical geometry.
In this setup, states with different topological orders may appear at 
different shifts $\mathcal{S}\equiv\frac{3}{2}N-N_\Phi$, where $N_\Phi$ is the 
number of flux quanta.

\begin{figure}[tb]
\begin{center}
\includegraphics[width=8.5cm]{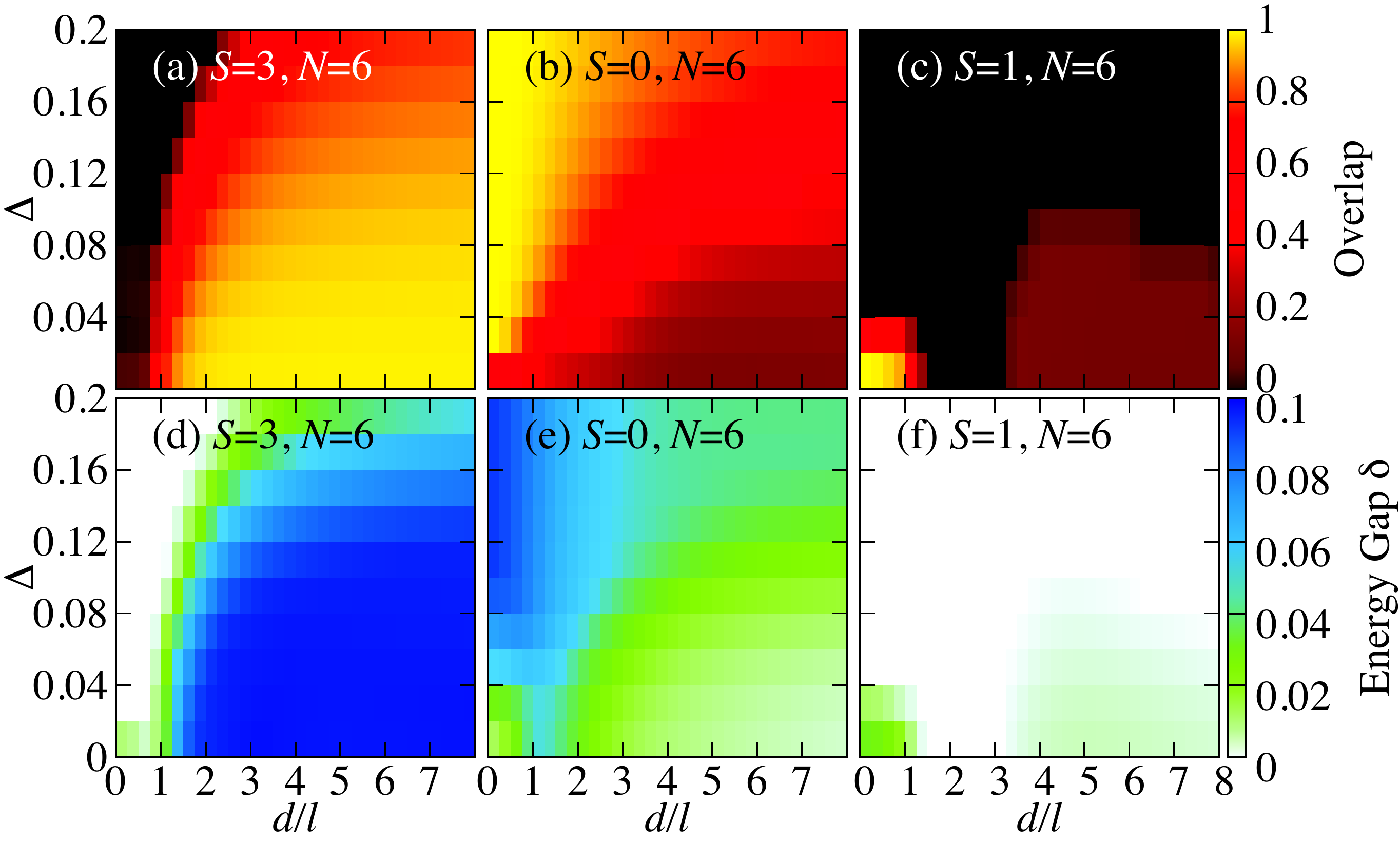}
\caption{Panels (a)-(f) correspond to shifts of $\mathcal{S}=3$, 0, and 1 from left to right for $N=6$ electrons in the lowest Landau level.  (a) - (c) show the overlap with $\Psi_{330}$, 
$2/3$ Laughlin, and the pseudo-spin singlet (112), respectively. (d) -(f) show the energy gap. 
 }  
\label{fig:over-gap-N6-LLL}
\end{center}
\end{figure}

\begin{figure}[tb]
\begin{center}
\includegraphics[width=8.5cm]{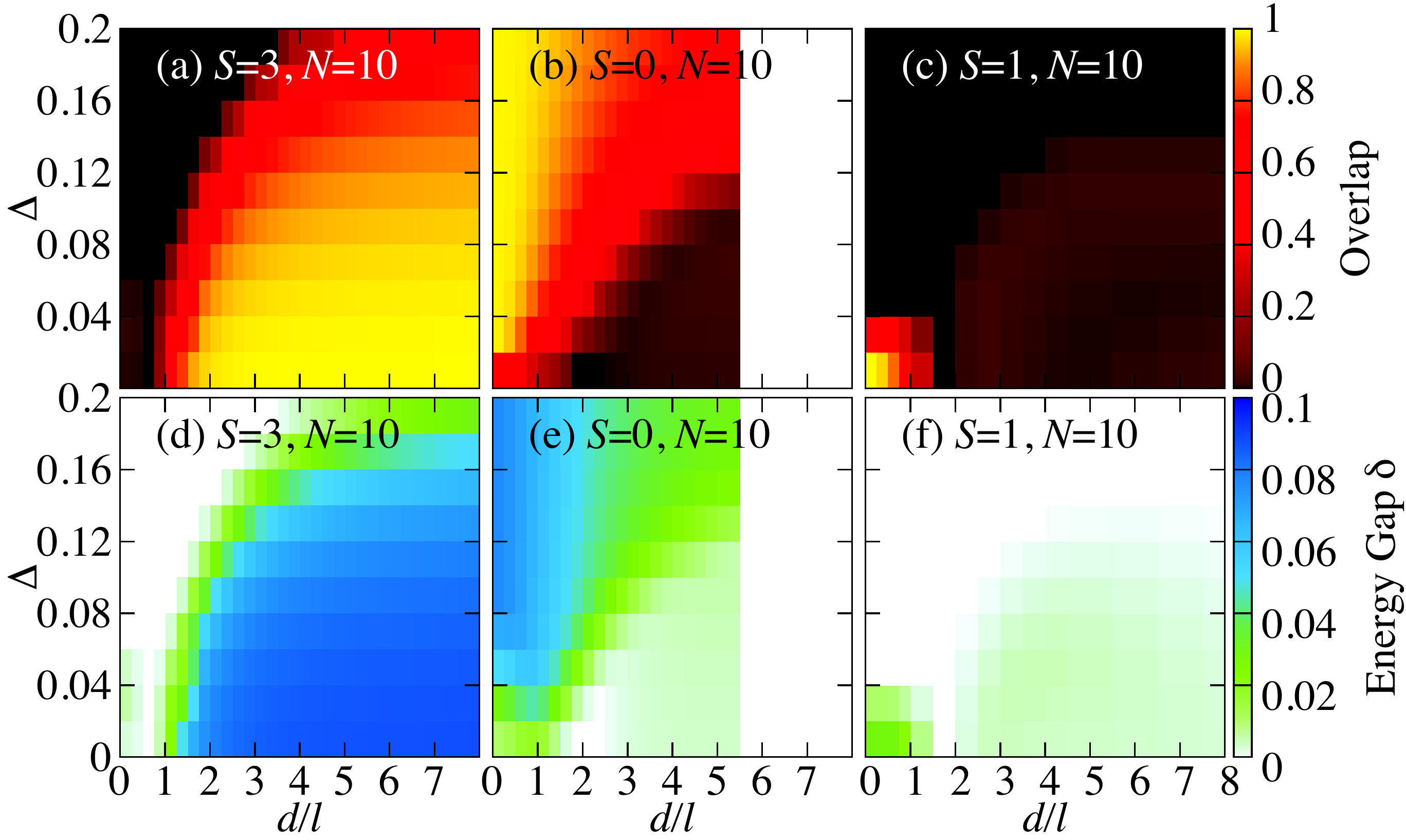}
\caption{Panels (a)-(f) correspond to shifts of $\mathcal{S}=3$, 0, and 1 from left to right for $N=10$ electrons in the lowest Landau level.  (a) - (c) show the overlap with $\Psi_{330}$, 
$2/3$ Laughlin, and the pseudo-spin singlet (112), respectively. (d) -(f) show the energy gap.  The white space beyond $d/l>5.5$ for $S=0$ was not calculated and therefore left blank.  
 }  
\label{fig:over-gap-N10-LLL}
\end{center}
\end{figure}

Figures \ref{fig:over-gap-N8-LLL}(a)-(f) displays our
numerical results for the overlaps of the model wave functions for the $330$, singlet $112$, and $2/3$-Laughlin states, 
with the exact Coulomb ground state at shifts $\mathcal{S}=3$, 1, and 0, respectively, together with the energy gaps for $N=8$. The energy gap is 
taken as the difference between the angular momentum $L = 0$ ground state and the first excited state (if the ground state has $L \neq 0$, the gap is set to zero).  
We can combine the energy gaps at different shifts into a single function 
$\delta(d/l,\Delta)$ by choosing the maximal gap among the different shifts.
We note this gap is not necessarily the transport gap  measured  experimentally but  the gap connected 
to the robustness of the phase--in many cases they are known to be qualitatively similar.  
Similar results are obtained for $N=6$ and 10 (see Figs.~\ref{fig:over-gap-N6-LLL} and~\ref{fig:over-gap-N10-LLL}).
We do not compute the overlap with the $Z_4$ parafermion state for $\mathcal{S}=3$ and $N=8$ for three reasons.  One is the $Z_4$ state is a single-component state and for $N=8$ electrons  there is only one possible $L=0$ state.  The second reason is the $Z_4$ state has four-electron clustering properties that cause it to vanish exactly unless $N\mod 4=0$.  Hence, one must consider at least $N=12$ electrons, see Figs.~\ref{fig:over-gap-N8-LLL}(g) and~\ref{fig:over-gap-N8-LLL}(h). 
The third reason is  the  gap at $\mathcal{S}=3$ in the single-component limit is significantly below the  gap at $\mathcal{S}=0$ corresponding to the 2/3 Laughlin state.  

We produce a phase diagram for the bilayer system taking into account both the 
overlaps and the energy gaps at different shifts.
The topological order is identified from the model wavefunction with the 
highest overlap with the ground state, and its stability is characterized by 
the energy gap.
Figure~\ref{fig:over-gap-N8-LLL}(i) shows a plot of the gap function 
$\delta(d/l,\Delta)$ for $N = 8$ and contour lines showing the wave function overlaps. Our results for $N=6, 10$ electrons (not shown) are 
consistent with Fig.~\ref{fig:over-gap-N8-LLL}(i). 
We emphasize  this approximate phase diagram matches remarkably well 
with  Ref.~\onlinecite{mcdonald1996}'s determined by wave function overlap and topological degeneracy on the torus.  

To investigate the relative stability of the $Z_4$ parafermion state, we consider 
the two-component system for $N = 12$ particles. 
In Figs.~\ref{fig:over-gap-N8-LLL}(g) and ~\ref{fig:over-gap-N8-LLL}(h), we show the overlap of the exact ground state with the $Z_4$ parafermion state, together with the value 
of the energy gap at shift $\mathcal{S} = 3$. While the $Z_4$ state has a maximum overlap
of $\approx 0.93$  in the single-component
limit, the Laughlin state has a much higher overlap 
of $\approx 0.99$. Furthermore, the system possesses a much larger energy gap at the 2/3 Laughlin shift  relative
to the shift of the $Z_4$ parafermion state.

\begin{figure}[t]
\begin{center}
  \includegraphics[width=7.cm,angle=0]{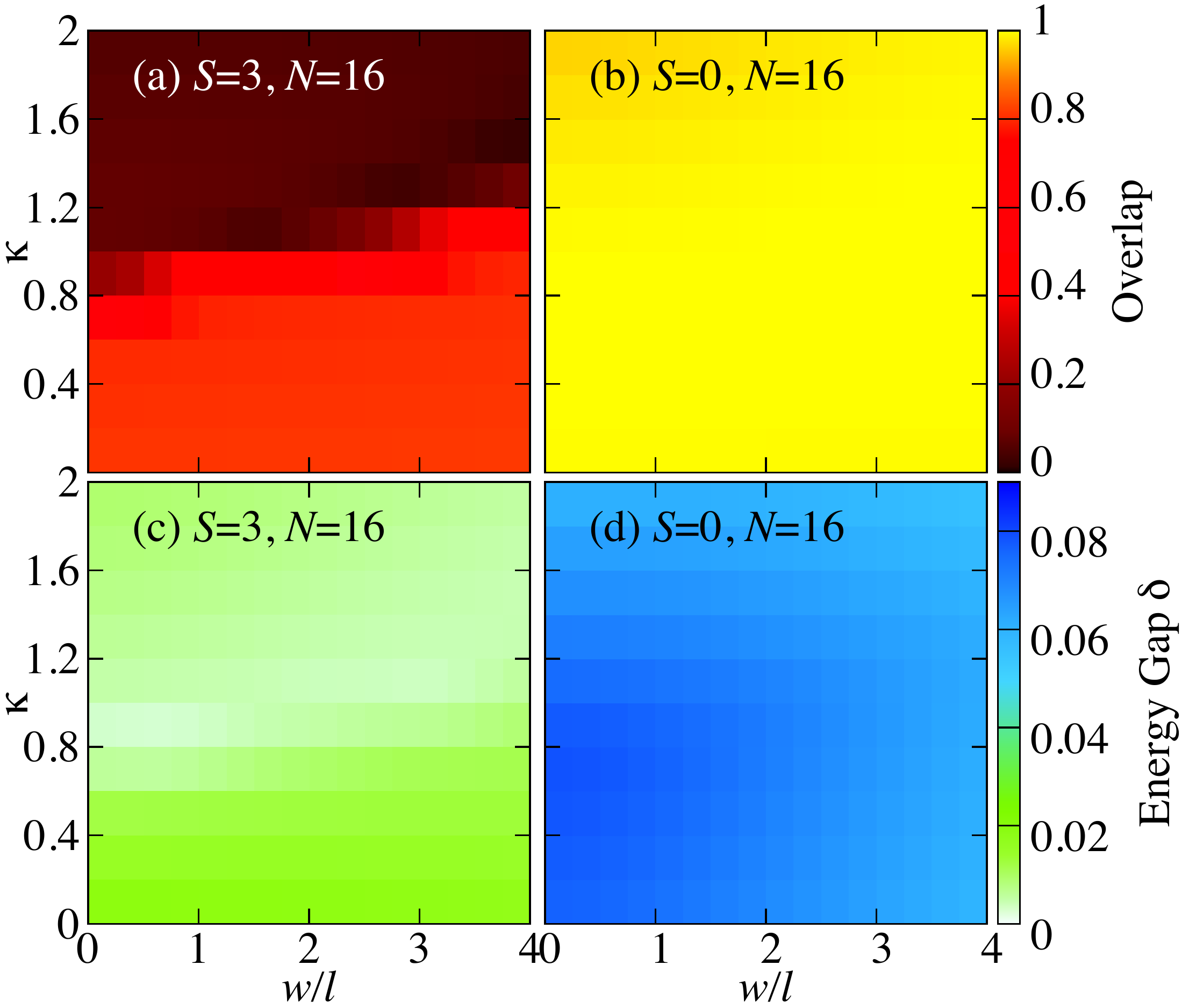}
  \caption{
\label{fig:N16-1C-LLL}
Overlap and gap calculations for $N = 16$ electrons in the LLL in the single-component limit. Top left (a) shows overlaps with the $Z_4$ state, at $\mathcal{S} = 3$; top right (b)
shows overlaps with the Laughlin state, at $\mathcal{S} = 0$. Lower panels, (c) and (d), show the gaps at $S = 3$ and $S = 0$.}
\end{center}
\end{figure}

To further assess the stability of the $Z_4$ state compared to the 2/3 Laughlin state, we can consider the single-component limit of Eq.~\eqref{Hbilayer}, 
obtained for strong tunneling $\Delta\gg 1$ and small $d/l\ll 1$, i.e., a single quantum well.
The smaller dimension of the Fock space in this limit allows us to consider $N=16$ electrons.
Here we study a particularly realistic model that includes Landau level (LL) mixing (parameterized by the ratio of the cyclotron energy to the Coulomb energy $\kappa=(\hbar\omega_c)/(e^2/\epsilon l)$) and finite width of the single quantum well (parametrized by well width $w/l$)~\cite{PhysRevB.87.245129}. 
Specifically, our realistic Hamiltonian is 
\begin{eqnarray}
H_\mathrm{realistic} &=& \sum_m V_m^{(2)}(w/l,\kappa)\sum_{i<j}P_{ij}(m)\nonumber\\
&+&\sum_m V_m^{(3)}(w/l,\kappa)\sum_{i<j<k}P_{ijk}(m)
\end{eqnarray}
where $P_{ij}(m)$ and $P_{ijk}(m)$ are operators that project onto pairs $(i,j)$ or triplets $(i,j,k)$ of electrons with relative angular momentum $m$.  The Hamiltonian 
is parameterized by two- and three-body pseudopoentials $V_m^{(2)}(w/l,\kappa)$ and $V_m^{(3)}(w/l,\kappa)$, respectively.  The two-body pseudopotentials are 
renormalized by Landau level mixing corrections which include the effects of virtual transitions of electrons (holes) to unoccupied (occupied) Landau and subband levels to lowest order in the Landau level mixing parameter $\kappa$.  Landau level mixing also produces an emergent three-body term that explicitly breaks particle-hole symmetry. 
This Hamiltonian is described in great detail in Ref.~\onlinecite{PhysRevB.87.245129} and was recently implemented in a numerical study of the FQHE at $\nu=5/2$~\cite{PhysRevX.5.021004}.
In this work we restrict our attention  to $w/l < 4$; wide quantum well systems are often better described as bilayers. 
Our results for the LLL are displayed in Fig.~\ref{fig:N16-1C-LLL} where the Laughlin state is clearly shown to be preferable. 
The $Z_4$ overlap at $\mathcal{S} = 3$ is large ($\approx 0.82$) and essentially decreases monotonically with $\kappa$ and 
is robust to width $w/l$. At $\mathcal{S} = 0$ the Laughlin state has an overlap of nearly unity ($\approx 0.99$) and  is robust to  
$\kappa$ and $w/l$. Both $\mathcal{S} = 0$ and $3$ have non-zero gaps, 
but the gap at $\mathcal{S} = 0$ is nearly three times larger than  $\mathcal{S}=3$.
Both overlaps and gaps are robust to varying $\kappa$ and $w/l$. 
Based on these results, we do not expect the $Z_4$ state in the bilayer system at $\nu = 2/3$ in the LLL.
Our conclusions for the LLL, based on ED, are further corroborated with variational Monte Carlo (see below).  

\begin{figure}[t]
\begin{center}
  \includegraphics[width=7.cm,angle=0]{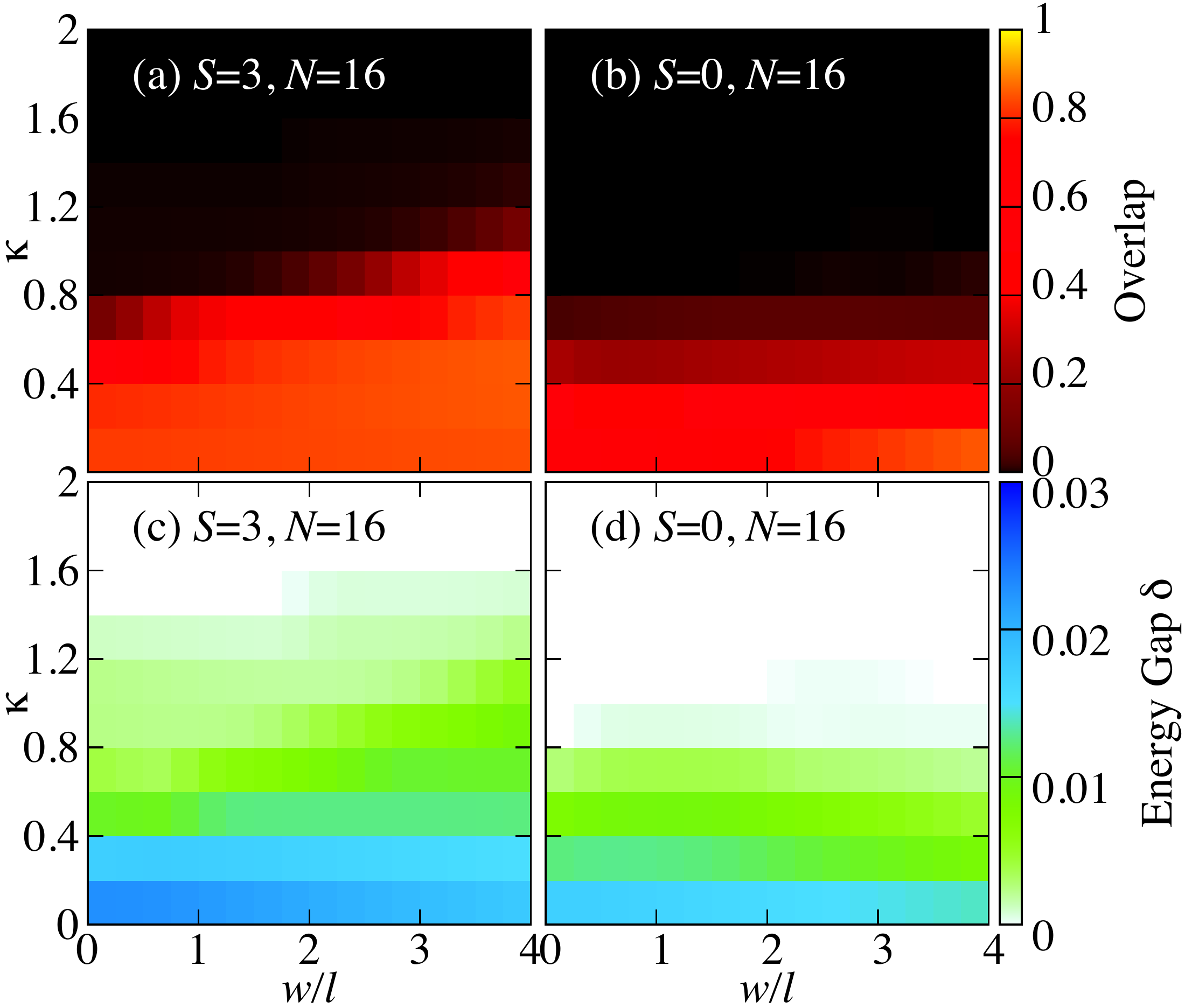}
  \caption{
\label{fig:N16-1C-SLL} Overlaps ((a) and (b)) and energy gaps ((c) and (d))  for $N = 16$ electrons in the SLL in the single-component limit.  See caption of Fig.~\ref{fig:N16-1C-LLL}.}
\end{center}
\end{figure}

\section{Second Landau Level}
Now that we have confidently ruled out 
the $Z_4$ state in the LLL, we turn our attention to the second Landau level (SLL).
Repeating the overlap and gap calculations with the SLL pseudopotentials, we 
obtain results that are quite different from the LLL.
In particular, we find in the single-component limit
($d/l \ll 1$ and $\Delta \gg 1$) the gap at shift $\mathcal{S} = 3$ is 
significantly larger than at $\mathcal{S}=0$ suggesting the ground state in 
this limit might not be the 2/3 Laughlin state, but rather an alternative 
state with $\mathcal{S}=3$.

Leaving the exploration of the full two-component phase diagram for the next 
section, we now take a closer look at the SLL in the single-component limit.
We use the same realistic model introduced earlier (but at filling $8/3$),
and focus on the competition between the $Z_4$ and the Laughlin states.
Surprisingly, we find that the $Z_4$ state 
appears favored over the Laughlin state according to both overlap 
and gap calculations, as shown in Fig. \ref{fig:N16-1C-SLL}. In the SLL, the overlap with the 
$Z_4$ state is qualitatively similar to the LLL, i.e., it is nearly $0.83-0.84$ for small $\kappa$ and decreases to zero as $\kappa$ is increased.
The Laughlin state at $\mathcal{S} = 0$ has a smaller overlap of $0.64- 0.8$, increases with $w/l$, and monotonically decreases with $\kappa$. 
Importantly, the gap is approximately 1.5 times larger at $\mathcal{S} = 3$ than it is at $\mathcal{S} = 0$.
Our results for $N = 12$ electrons are qualitatively similar, but with quantitatively higher overlaps
for both the $Z_4$ and Laughlin states. 
Last, we note that in the limit of zero LL mixing ($\kappa=0$) our Hamiltonian is particle-hole symmetric and our results for the spin-polarized $\nu=8/3$ state should translate to $\nu=7/3$ where some recent theoretical studies have suggested that the FQH state at 7/3 is likely in the Laughlin universality class~\cite{JohriPRB2014,ZaletelPRB2015}.

\begin{figure}[t]
\begin{center}
\includegraphics[]{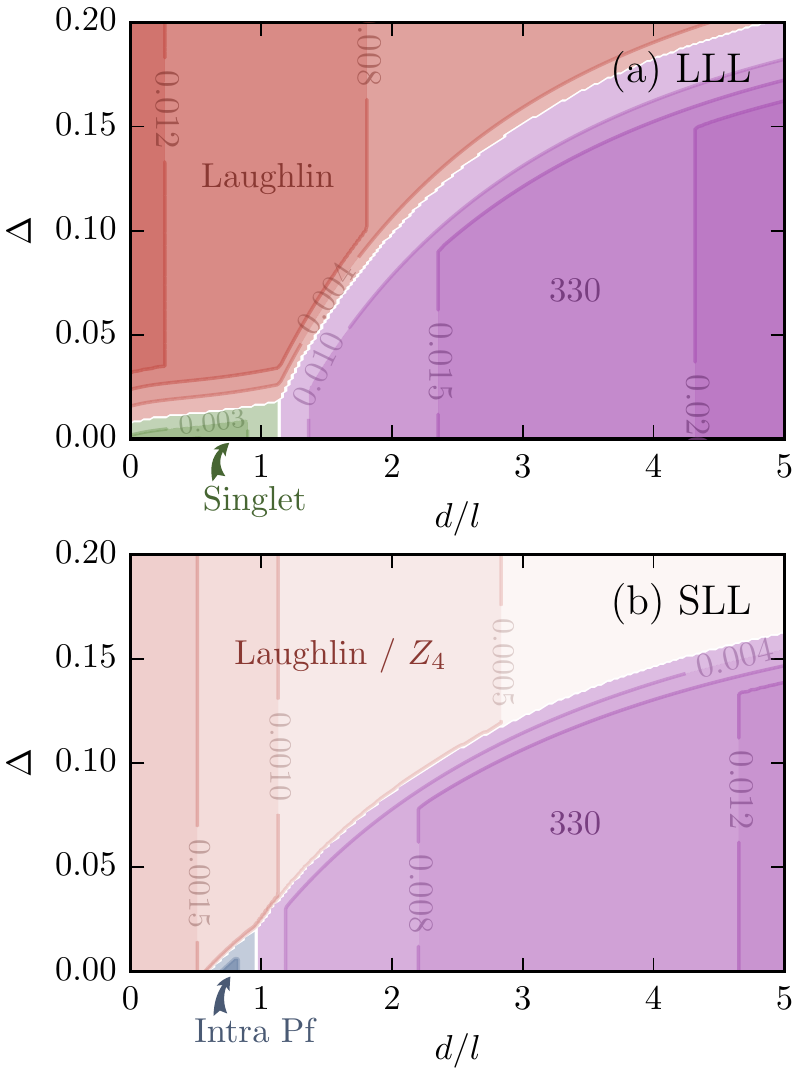}
\caption{The quantum phase diagram determined by variational energies at 
thickness $w=0$.
The contour lines depict the energy advantage $\delta E$ of each dominant phase.
In the SLL, the Laughlin state does not have a clear energy advantage over the 
$Z_4$ state.
}  
\label{fig:MC-LLL-SLL}
\end{center}
\end{figure}

\section{Variational Energies}
The discussion so far has focused on the fate of the $Z_4$ state in comparison 
with the three Abelian phases, without exploring other non-Abelian 
possibilities.
One may also worry about the finite-size effect in the ED results.
To address these concerns, we have performed variational energy calculations
that also include the interlayer and intralayer Pfaffian states at shift 
$\mathcal{S}=3$, for much larger system sizes.
The energy expectation value of Eq.(\ref{Hbilayer}) of the three Abelian states and the Pfaffians are computed using 
Monte Carlo for up to $N=60$ electrons, with sample size $10^7$.
We refer to Ref.~\onlinecite{Morf87} for the details of the Monte Carlo energy
calculation and Ref.~\onlinecite{DavenportPRB2012} for the efficient 
evaluation of composite fermion wavefunctions.
It turns out that the anti-symmetrization used to construct $\Psi_{Z_4}$ [Eq.~(\ref{eq:z4})] is prohibitively expensive for numerical calculations, hence, we leverage the Jack polynomial representation of $\Psi_{Z_4}$ to directly obtain its second-quantized amplitudes~\cite{JackGenerator}.  This technique allows us to obtain the  state and compute its variational energy, i.e., the expectation value of Hamiltonian for $\Psi_{Z_4}$, for up to $N=28$ electrons, well beyond the scope of exact diagonalization.
We assess the relative stability of different phases of the two-component 
system by comparing their variational energies.
To estimate the energy per particle in the thermodynamic limit, we use 
quadratic extrapolation in $1/N$, weighted by the statistical error on each 
data point~\cite{DavenportPRB2012}.
The extrapolated energy has an error between $10^{-3}$ and $10^{-4}$.

The phase diagram is determined according to the wavefunction with the 
lowest energy, and we characterize the phase stability using the energy advantage 
$\delta E$ of the dominant wavefunction over its closest competitor.
Figure~\ref{fig:MC-LLL-SLL} shows the contour plots of $\delta E(d/l,\Delta)$.
In the LLL, the phase diagram determined by variational energies 
is in qualitative agreement with the ED results.
We find that the non-Abelian $Z_4$ and the interlayer/intralayer Pfaffian 
states remain energetically unfavorable throughout the phase diagram,
and the singlet 112 occupies a very small corner of the parameter space.
In the SLL, while the Halperin 330 state still dominates at large layer 
separation $d/l$, the Laughlin state at large $\Delta$ is now much less 
stable compared with the LLL.
The main competition comes from the non-Abelian $Z_4$ state.
In fact, for much of the phase diagram, the energy difference between the 
two is on the same order as the estimated extrapolation error 
$(\lesssim 10^{-3})$ for the $Z_4$ state.
This is in strong agreement with our gap and overlap calculations using ED, 
namely, that the non-Abelian $Z_4$ state is highly competitive with the 
Laughlin state.
Incidentally, we also find a small region in the parameter space that favors 
the intralayer Pfaffian state, but we do not find any parameter set that 
stabilizes the interlayer Pfaffian state.

\section{Conclusion}
Based on our exact diagonalization and Monte Carlo studies, we find that 
$\nu = 2/3$ bilayers in the LLL, in the limit of weak LL mixing, most likely do not realize the 
non-Abelian $Z_4$ parafermion state. 
Most remarkably, in the single-component limit of the SLL,  the non-Abelian $Z_4$ phase may be favorable for the $8/3$ FQH state relative to the Laughlin state. 
Indeed, previous  studies of the experimentally obtained energy gaps of FQH states in the SLL have already indicated the possibility that the electron correlations are sufficiently different from those of the LLL and that novel exotic states might be realized~\cite{choi2008,PanPRL1999,PanPRB2008,dean2008,KumarPRL2010,SamkharadzePRB2011,Kleinbaum2015}.
While weak quasiparticle tunneling experiments through a quantum point contact~\cite{BaerPRB2014} suggest that $\nu=8/3$ is the 2/3 Laughlin state, it cannot be considered to be definitive yet and more experiments are necessary.
The implication of our finding that the 
observed $8/3$ SLL FQH state may be the parafermionic $Z_4$ non-Abelian phase is enormous since this 
 state can be utilized for universal topological  quantum computation.
  
 As this work was being completed, we became aware of a related manuscript~\cite{PhysRevB.91.205139} by Geraedts et al. By utilizing primarily the density matrix renormalization group technique, they reported that the interlayer Pfaffian is stabilized for a modified LLL interaction with a hollow core, which is very different from the LLL and SLL realistic Coulomb interactions considered in our work.  It is an interesting open question whether the disparity between the present work and that of Ref.~\onlinecite{PhysRevB.91.205139} is due to differences in techniques, differences in models, or both.

\begin{acknowledgements}
M.R.P. thanks the Office of Research and Sponsored Programs at California State University Long Beach and Microsoft Station Q. We thank M. Zaletel, P. Bonderson,  and N. Regnault  for useful discussions.   Additionally we thank G. Cs\'athy and S. Davenport for help comments on the manuscript.  
\end{acknowledgements}


\end{document}